\documentclass[a4paper,11pt]{article}
\usepackage{pos}
\usepackage{float}
\usepackage{xspace}

\newcommand{\aspi}{\ensuremath{a_S}}
\newcommand{\fbinv}{\ensuremath{\,\mathrm{fb}^{-1}}}
\newcommand{\GeV}{\ensuremath{\,\text{Ge\hspace{-.08em}V}}}
\newcommand{\muf}{\ensuremath{\mu_f}}
\newcommand{\mur}{\ensuremath{\mu_r}}
\newcommand{\MSbar}{\ensuremath{\overline{\textrm{MS}}}}
\newcommand{\pp}{\ensuremath{\mathrm{p}{}\mathrm{p}{}}}
\newcommand{\PQt}{\ensuremath{\mathrm{t}}}
\newcommand{\MCFM} {\textsc{MCFM}\xspace}
\newcommand{\mt}{\ensuremath{m_{\PQt}}}
\newcommand{\mtpole}{\ensuremath{m_{\PQt}^{\mathrm{pole}}}}
\newcommand{\mtMSbar}{\ensuremath{\overline{m}_{\PQt}}}
\newcommand{\mtMSR}{\ensuremath{m_{\PQt}^{\mathrm{MSR}}}}
\newcommand{\TeV}{\ensuremath{\,\text{Te\hspace{-.08em}V}}}
\newcommand{\ttbar}{\ensuremath{{\PQt{}\overline{\PQt}}}}

\title{Single-differential top quark pair production cross sections with running mass schemes at NLO}
\ShortTitle{Single-differential $t\overline{t}$ production cross sections with running mass schemes at NLO}

\author*[a,b]{Toni M{\"a}kel{\"a}}
\author[c]{Andr{\'e} Hoang}
\author[a,d]{Katerina Lipka}
\author[e]{Sven-Olaf Moch}

\affiliation[a]{Deutsches Elektronen-Synchrotron,\\
                Notkestraße 85, D-22607 Hamburg, Germany}

\affiliation[b]{National Centre for Nuclear Research,\\
                Pasteura 7, 02-093 Warsaw, Poland}
                
\affiliation[c]{Fakult{\"a}t f{\"u}r Physik, 
                Universit{\"a}t Wien,\\
                Boltzmanngasse 5, A-1090 Wien, Austria}

\affiliation[d]{Fakult{\"a}t f{\"u}r Mathematik und Naturwissenschaften,  
                Bergische Universit{\"a}t Wuppertal,\\
                Gaussstrassse 20, 42119 Wuppertal, Germany}

\affiliation[e]{II. Institut für Theoretische Physik, 
                Universit{\"a}t Hamburg,\\
                Luruper Chaussee 149, 22761 Hamburg, Germany}

\emailAdd{Toni.Makela@ncbj.gov.pl}

\abstract{Single-differential cross section predictions for top quark pair production are presented at next-to-leading order, using running top quark mass renormalization schemes. The evolution of the mass of the top quark is performed in the MSR scheme $\mtMSR(\mu)$ for renormalization scales $\mu$ below the $\MSbar$ top quark mass $\mtMSbar(\mtMSbar)$, and in the $\MSbar$ scheme $\mtMSbar(\mu)$ for scales above. In particular, the implementation of a mass renormalization scale independent of the strong coupling renormalization scale and factorization scale in quantum chromodynamics allows investigating independent dynamical scale variations. Furthermore, the first theoretically consistent extraction of the top quark MSR mass from experimental data is presented.}

\FullConference{%
  41st International Conference on High Energy physics - ICHEP2022\\
  6-13 July, 2022\\
  Bologna, Italy
}

\begin{document}
\maketitle

The top quark mass $\mt$ is a fundamental parameter of the Standard Model and has an important role in many predictions, both directly and via higher-order corrections. 
Yet, the formal definition of quark masses makes them renormalization scheme dependent quantities. While the pole mass $\mtpole$ suffers from the renormalon ambiguity, an infrared sensitivity of the order of the scale of quantum chromodynamics (QCD)~\cite{Bigi:1994em}, short-distance masses e.g. the $\MSbar$ mass $\mtMSbar(\mu_m)$ and the MSR mass $\mtMSR(R)$~\cite{Hoang:2017btd} do not. However, the dependence on the mass renormalization scales $\mu_m$ and $R$ necessitates proper scale setting for the extraction of theoretically well-defined masses from cross section measurements to avoid the appearance of large logarithms.

\begin{figure}[ht]
\centering
\includegraphics[width=0.4238\textwidth]{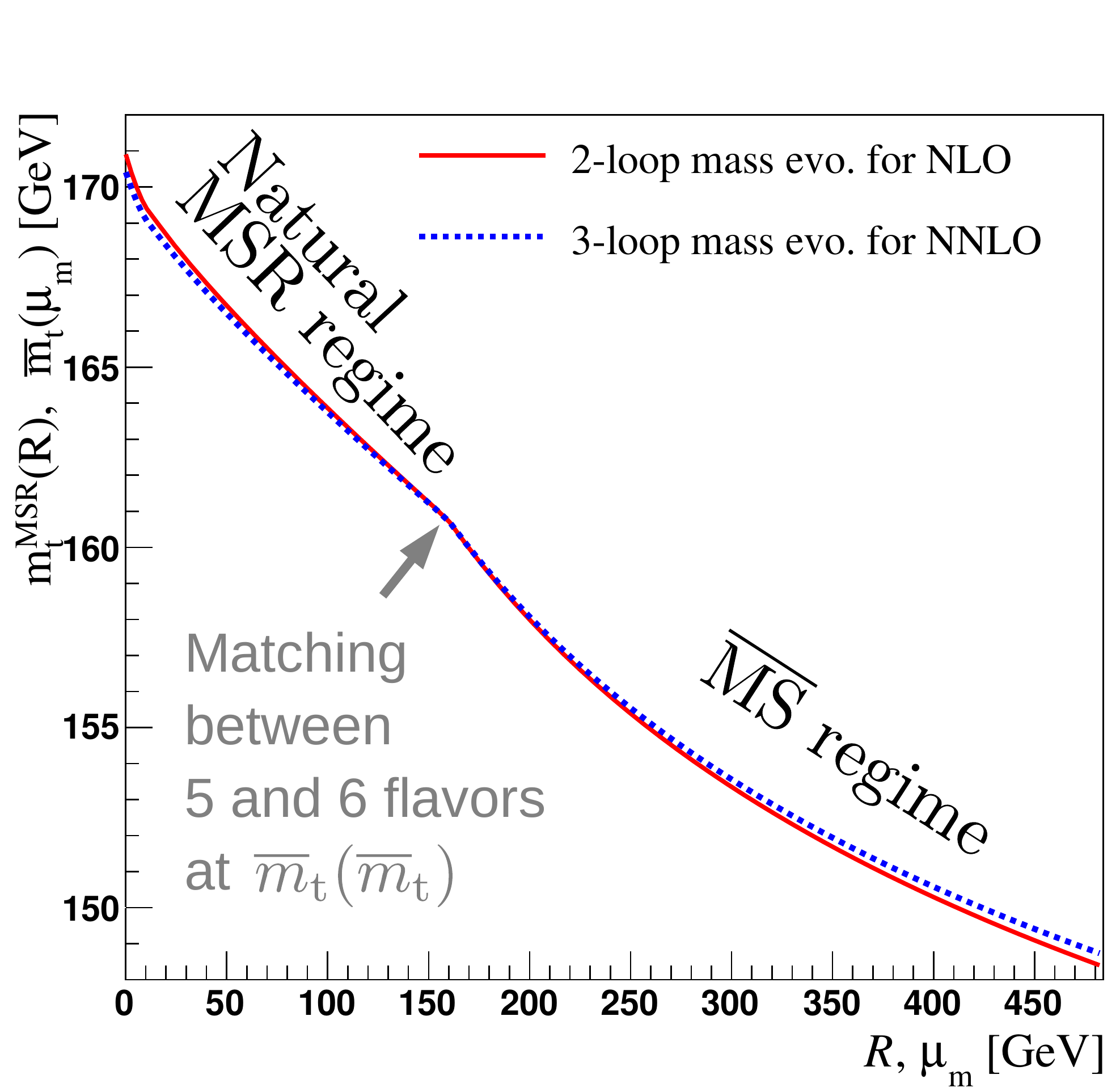}
\hspace*{1cm}
\raisebox{0.9mm}{\includegraphics[width=0.391\textwidth]{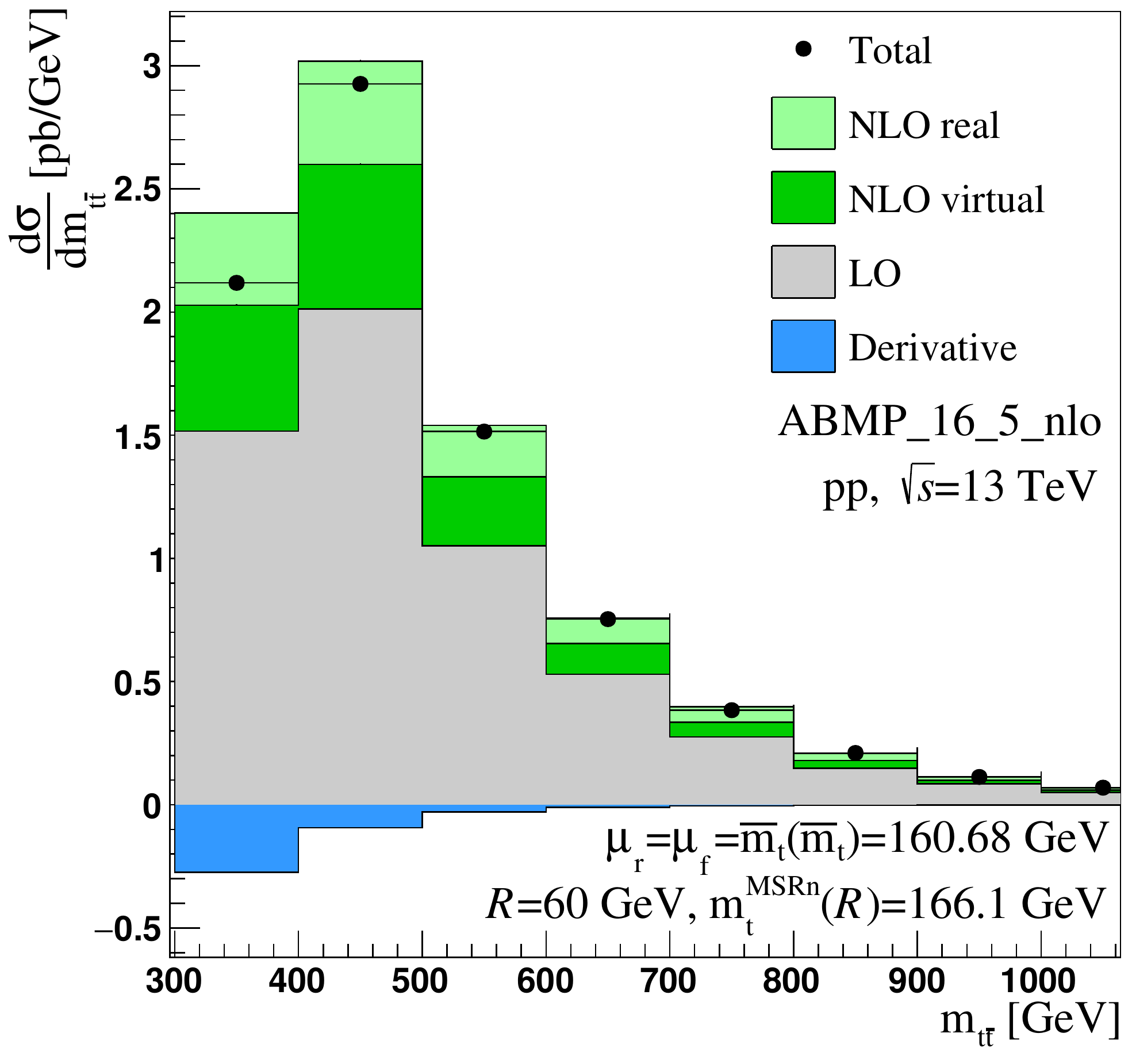}}
\caption{
\textit{Left}: the running of $\mt$ in the MSR and $\MSbar$ evolution regimes, see also Ref.~\cite{Hoang:2017btd}.
\textit{Right}: the $\ttbar$ production cross section at NLO in bins of $m_\ttbar$ (dots) and the contributions of the terms in Eq.~\eqref{diff_NLO_cs} (histograms).
}
\label{mEvo_m34_R60}
\end{figure}

The pole and $\MSbar$ masses are related by
$\mtpole = \mtMSbar(\mu_m)
           \left( 1 + \sum_{n=1} d^{\MSbar}_n(\mu_m) (\aspi(\mu_m))^n \right)$,
where $\aspi \equiv \alpha_S/\pi$, 
and $d^{\MSbar}_n(\mu_m)$ are perturbative coefficients.
The pole and MSR mass relation reads
$\mtpole = \mtMSR(R) + R \sum_{n=1}^{\infty} d^{\textrm{MSR}}_n (\aspi(R))^n$,
i.e. $\mtMSR(R)$ approaches $\mtpole$ in the formal limit $R\rightarrow 0$, and the $\MSbar$ mass at $R\rightarrow \mtMSbar(\mtMSbar)$ up to a small matching correction. The latter are obtained by integrating out top quark loop corrections at $R \lesssim \mtMSbar(\mtMSbar)$~\cite{Hoang:2017suc}.
The $R$-evolution of $\mtMSR(R)$ is linear, contrary to the logarithmic $\mu_m$ evolution of $\mtMSbar(\mu_m)$, and designed to capture the correct physical logarithms for observables with $\mt$ dependence generated at dynamical scales $R<\mt$ (e.g. resonances, thresholds, low-energy endpoints)~\cite{Hoang:2021fhn}. 
For dynamical scales of order and larger than $\mt$, the $\MSbar$ mass and evolution are used. 
Based on Ref.~\cite{Dowling:2013baa}, the top quark-antiquark ($\ttbar$) production cross section as a function of the $\ttbar$ system invariant mass $m_\ttbar$ at next-to-leading order (NLO) reads
\begin{equation}
\frac{d\sigma}{dm_\ttbar}
=  \left( \aspi \right)^2
   \frac{d\sigma^{(0)}}{dm_\ttbar}\big(m,\mur,\muf\big)
 + \left( \aspi \right)^3
   \frac{d\sigma^{(1)}}{dm_\ttbar}\big(m,\mur,\muf\big)
 + \left( \aspi \right)^3
   \tilde{R} d_1
   \frac{d}{d \mt}\left(\frac{d\sigma^{(0)}(\mt,\mur,\muf)}{dm_\ttbar}\right)
                 \bigg|_{\mt = m},
\label{diff_NLO_cs}
\end{equation}
where $\sigma^{(0)}$ is the leading order (LO) and $\sigma^{(1)}$ the NLO cross section in the pole mass scheme and the derivative term at NLO implements the $\MSbar$ or MSR top mass schemes, the renormalization (factorization) scale $\mur$ ($\muf$) is independent of the mass renormalization scales $R$ or $\mu_m$ and $\aspi = \aspi(\mur)$. Furthermore,
\begin{equation}
(m, d_1, \tilde{R})
= 
\begin{cases}
(\mtMSR(R),~d^{\textrm{MSR}}_1,~R),
& \textrm{in the MSR regime }(R<\mtMSbar(\mtMSbar)),\\
(\mtMSbar(\mu_m),~d^{\MSbar}_1(\mu_m),~\mtMSbar(\mu_m)),
& \textrm{in the $\MSbar$ regime }(R>\mtMSbar(\mtMSbar)).
\end{cases}
\end{equation}
In this work, the cross section given in Eq.~\eqref{diff_NLO_cs} is implemented into \MCFM v6.8~\cite{Campbell:2010ff}. 
The running of $\mt$ and the $\ttbar$ cross section as a function of $m_\ttbar$ are shown in Fig.~\ref{mEvo_m34_R60}.

\begin{figure}[ht]
\centering
\includegraphics[width=0.43\textwidth]{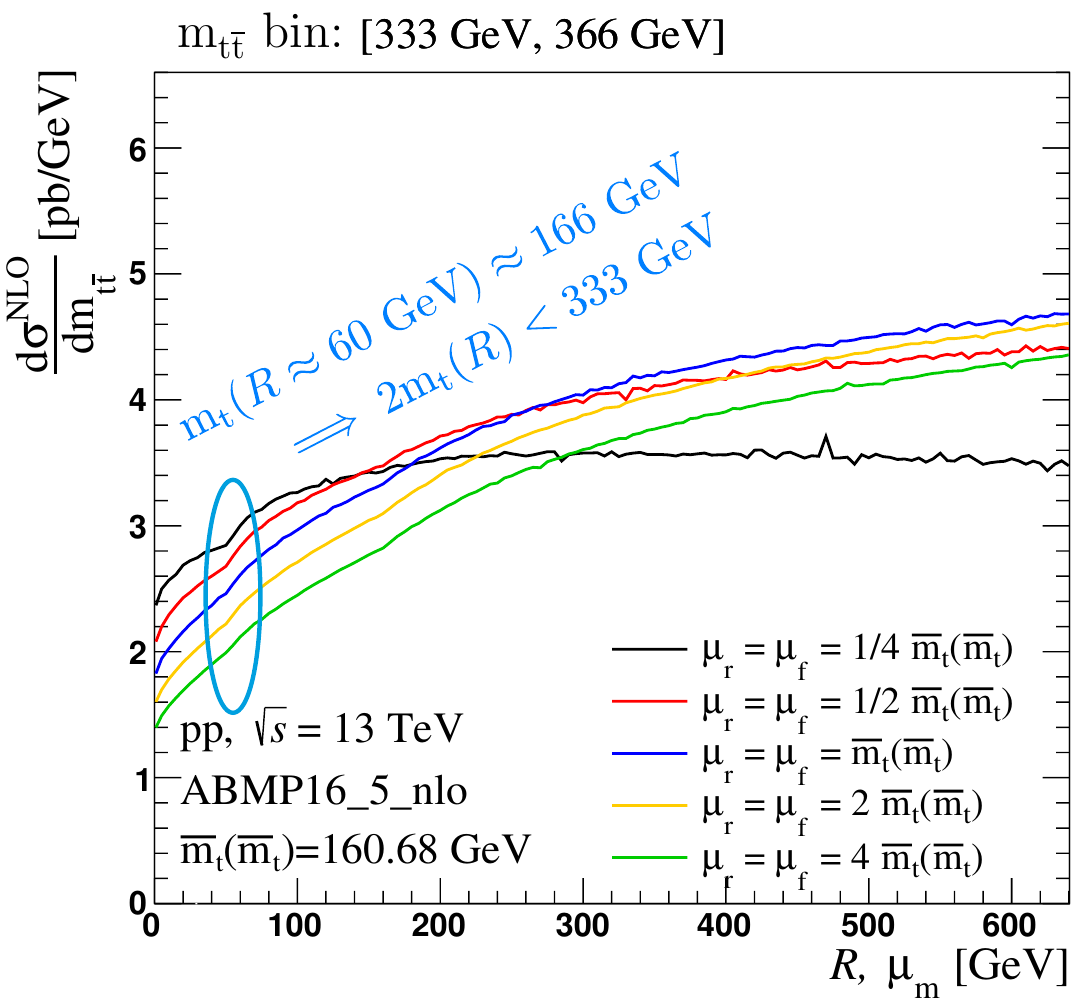}
\caption{The $m_\ttbar \in [333,366]\GeV$ bin of $d\sigma/dm_\ttbar$ as a function of $R,~\mu_m$. At $R\lesssim60\GeV$, threshold effects are prominent. Low values of $\mur,~\muf$ are observed to stabilize the predictions as a function of $R,~\mu_m$.}
\label{m34_333-366}
\end{figure}

The implementation of the mass renormalization scales independently from $\mur$ and $\muf$ allows the first investigation of the dependence on the scale $R$. 
As illustrated in Fig.~\ref{m34_333-366} for $d\sigma/dm_\ttbar$ cross section in the bin $m_\ttbar \in [333,366]\GeV$, low values of $\mur$ and $\muf$ result in quick stabilization of the NLO $\ttbar$ cross section as a function of the mass renormalization scale in this bin, which contains high sensitivity to $\mt$.
Furthermore, Fig.~\ref{m34_333-366} indicates unsmooth behavior at low $R$, and the cross section stabilizes at $R \gtrsim 60\GeV$. This is attributed to Coulomb effects spoiling the convergence of expansions in $\alpha_S$ in fixed-order QCD near the $\ttbar$ production threshold. In future studies, this is to be solved by including quasi-boundstate corrections and the resummation of soft gluon effects, following e.g. Ref.~\cite{Kiyo:2008bv}.
In accordance with these observations, the central values of $R$, $\mur$ and $\muf$ are set to $80\GeV$ in the following to obtain predictions robust against scale variations.

A determination of $\mtMSR(R)$ is performed using the single-differential $\ttbar$ production cross section measured by the CMS Collaboration in $\pp$ collisions at $\sqrt{s}=13\TeV$~\cite{CMS:2019jul}, corresponding to an integrated luminosity of $35.9 \fbinv$. The cross section is provided in four bins: 
$m_\ttbar < 420\GeV$, 
$m_\ttbar \in [420, 550]\GeV$, 
$m_\ttbar \in [550, 810]\GeV$ and
$m_\ttbar > 810\GeV$.
The top quark MSR mass is extracted by fitting $\ttbar$ production cross section predictions, computed with the ABMP16 5 flavor PDF~\cite{Alekhin:2017kpj} at NLO, to the experimental data.
For the extraction, $R=80\GeV$ is assumed. For comparison with previous studies, the resulting $\mtMSR(80\GeV)$ is evolved to the reference scale $R=1\GeV$, as well as translated to $\mtMSbar(\mtMSbar)$. The fit uncertainty is obtained via the $\Delta\chi^2 = 1$ tolerance criterion. The uncertainty in the initial $R$ choice is estimated by repeating the fits at $R=60\GeV$ and $100\GeV$, and taking the difference of the masses evolved to the reference scales to the respective results of the $R=80\GeV$ fit. The $\mur$, $\muf$ uncertainty is obtained by multiplying the scales independently by $2^{\pm 1}$, avoiding cases where one scale is multiplied by 2 and the other by 1/2, and constructing an envelope.

With $\mur=\muf=\mtMSR(80\GeV)$ throughout the $m_\ttbar$ distribution, evolving the obtained $\mtMSR(80\GeV)$ to $R=1\GeV$ yields 
$\mtMSR(1\GeV) = 173.2 \pm 0.6\,\text{(fit)}
                       ^{+0.4}_{-0.6}\,\text{($\mur,\muf$)}
                       ^{+0.4}_{-0.5}\,\text{($R$)} \GeV$,
compatible with the Monte Carlo calibration studies performed in Ref.~\cite{ATLAS:2021urs} where $R=1\GeV$ was also adopted as the reference scale. 
Furthermore, the result translates into 
$\mtMSbar(\mtMSbar) = 163.3^{+0.8}_{-1.0} \GeV$, 
which is in agreement with Ref.~\cite{CMS:2019jul}. 
However, it disagrees with Ref.~\cite{CMS:2019esx}, where 
$\mtpole = 170.5 \pm 0.8 \GeV$ 
was obtained, which translates into 
$\mtMSR(1 \GeV) = 170.2 \pm 0.8 \GeV$ 
interpreting the pole mass~\cite{CMS:2019esx} as the asymptotic pole mass~\cite{Hoang:2021fhn}.

Setting $\mur$ and $\muf$ to $\mtMSR(80\GeV)/2$ for $m_\ttbar < 420\GeV$ and to $\mtMSR(80\GeV)$ for $m_\ttbar > 420\GeV$ yields 
$\mtMSR(1\GeV) = 174.8 \pm 0.5\,\text{(fit)}
                        ^{+0.2}_{-0.4}\,\text{($\mur,\muf$)}
                        ^{+0.2}_{-0.3}\,\text{($R$)} \GeV$.
As expected from the present investigations, the setting increases robustness against scale variations, resulting in small uncertainties. 
Though a full understanding of $\mt$ extracted from cross section measurements requires the inclusion of threshold Coulomb effects, the results indicate
that the choice of the top quark mass scheme and the use of dynamical renormalization scales have a considerable impact on the phenomenological analysis and need to be investigated thoroughly.

\end{document}